\documentclass[]{aastex631}

\begin{document}

\title{Interaction of the Prominence Plasma within the Magnetic Cloud of an ICME with the Earth’s Bow Shock}
\correspondingauthor{Hadi Madanian}
\email{hadi.madanian@nasa.gov}
%
%

\author[0000-0002-2234-5312]{Hadi Madanian}
\affiliation{NASA Goddard Space Flight Center, Greenbelt, MD 20771, USA}
\affiliation{Catholic University of America, Washington, DC 20064, USA}

\author{Li-Jen Chen}
\affiliation{NASA Goddard Space Flight Center, Greenbelt, MD 20771, USA}

\author{Jonathan Ng}
\affiliation{NASA Goddard Space Flight Center, Greenbelt, MD 20771, USA}
\affiliation{University of Maryland, College Park, MD 20742, USA}

\author{Michael J. Starkey}
\affiliation{Southwest Research Institute, San Antonio, TX 78238, USA}

\author{Stephen A. Fuselier}
\affiliation{Southwest Research Institute, San Antonio, TX 78238, USA}

\author{Naoki Bessho}
\affiliation{NASA Goddard Space Flight Center, Greenbelt, MD 20771, USA}
\affiliation{University of Maryland, College Park, MD 20742, USA}

\author{Daniel J. Gershman}
\affiliation{NASA Goddard Space Flight Center, Greenbelt, MD 20771, USA}

\author{Terry Z. Liu}
\affiliation{University of California Los Angeles, Los Angeles, CA 90095, USA}



\begin{abstract}
The magnetic cloud within an interplanetary coronal mass ejection (ICME) is characterized by high magnetic field intensities. In this study, we investigate the interaction of a magnetic cloud carrying a density structure with the Earth's bow shock during the ICME event on 24 April 2023. Elevated abundances of cold protons and heavier ions, namely alpha particles and singly charged helium ions, associated with the prominence plasma are observed within this structure. The plasma downstream of the bow shock exhibits an irregular compression pattern which could be due to the presence of heavy ions. Heavy ions carry a significant fraction of the upstream flow energy; however, due to their different charge per mass ratio and rigidity, they are less scattered by the electromagnetic and electrostatic waves at the shock. We find that downstream of the shock, while the thermal ion energy is only a small fraction of the background magnetic energy density, nevertheless increased ion fluxes reduce the characteristic wave speeds in the that region. As such, we observe a transition state of an unstable bow shock layer across which the plasma flow is super Alfvénic in both upstream and downstream regions. Our findings help with understanding the intense space weather impacts of such events.
\end{abstract}



\section{Introduction}

At the Earth’s bow shock, protons are the most abundant ion species in the solar wind plasma and determine the scales of the main shock processes. The typical solar wind plasma also contains alpha particles (He\textsuperscript{++}, $\sim\%$4), and trace abundances of other ion species in various charge states (He\textsuperscript{+}, O\textsuperscript{7+}, O\textsuperscript{6+}, Fe\textsuperscript{6+}, etc.) \citep{bame_solar_1968}. During coronal mass ejections (CMEs), large quantities of charged particles are released in eruptive events at the Sun and are accelerated through the solar corona into the interplanetary medium. The eruption drives a shock wave ahead of the interplanetary CME (ICME) which heats and compresses the solar wind particles and creates a sheath of hot plasma ahead of the ICME \citep{wimmer-schweingruber_understanding_2006}. The plasma behind the sheath can carry a magnetic flux rope known as the magnetic cloud characterized by a strong magnetic field and a tenuous plasma (i.e., high Alfvén speed ($v_{Alf}$) and low ion $\beta$). At times, prominence materials of filamentary structures at the Sun’s chromosphere can find their way into the magnetic cloud without experiencing much coronal heating \citep{wang_cold_2018,skoug_prolonged_1999, burlaga_magnetic_1998}, resulting in transient enhancements in the abundances of cold, low charge state heavy ions, especially He\textsuperscript{+}, in the solar wind \citep{cane_interplanetary_1986, gosling_observations_1980,schwenn_singlyionized_1980}. Rare instances of observing enhanced fluxes of cold heavy ions near the Earth's bow shock are unique opportunities to study their heating process and their impact on collisionless shock waves.

Collisionless shocks are the boundary between two states of plasma with different energy distributions. Across the Earth's bow shock, the upstream kinetic energy is dissipated and transferred to heating and thermalization of charged particles. Several dispersive and dissipative mechanisms contribute to ion heating at collisionless shocks \citep{kennel_quarter_1985, sckopke_ion_1995, krasnoselskikh_dynamic_2013}. A shock wave is characterized by its Alfvénic Mach number ($M_A$) defined as the ratio of the shock speed to the upstream Alfvén speed $v_{Alf}$. Supercritical shocks ($M_{A} > \sim 3$) reflect ions to dissipate energy \citep{treumann_fundamentals_2009}. At marginally critical shocks, ion heating includes the entire bulk ion distribution and occurs entirely within the shock ramp \citep{thomsen_ion_1985, wilkinson_ion_1991}. At high Mach number supercritical shocks, where ion reflection from the shock is significant \citep{madanian_drivers_2024}, reflected ions play a major role in dissipating energy and carrying the enthalpy flux downstream of the shock \citep{schwartz_energy_2022, fuselier_h_1994}. At these shocks, while heating and deceleration of the solar wind begin in the foot region, the core population is mainly slowed down within the shock ramp where electromagnetic waves contribute significantly to perpendicular heating of the core ion population and parallel electron acceleration \citep{agapitov_energy_2023, wilson_observations_2012}. These waves are typically in the whistler mode frequency range which can be emitted by the shock or generated by instabilities associated with cross-field currents or temperature anisotropies  \citep{burgess_microstructure_2016, tidman_emission_1968}. 

The angle between the shock wave propagation direction and the magnetic field $\theta_{Bn}$ is typically used to classify shocks as quasi-parallel ($\theta_{Bn} < 45^{\circ}$) or quasi-perpendicular ($\theta_{Bn} > 45^{\circ}$). At quasi-parallel shocks perturbations and heating processes can begin at large distances upstream of the shock whereas in quasi-perpendicular shocks most upstream waves are limited to a short ion gyroradius distance.

Analytical models of 1D thin shock layers suggest that gyration of ions in the downstream at different phases can collectively cause ion heating and temperature anisotropies \citep{lee_heating_2000,gedalin_ion_1997}. Theoretically, Liouville mapping of the Hamiltonian version of the ion equations of motion across the shock also results in downstream temperature anisotropy \citep{ellacott_heating_2003}. The simplified nature of these models gives abundance of versatility to study different shock problems with static fields but at the expense of losing physical mechanisms involving waves and microinstabilities. 2D hybrid simulations on the other hand have shown a complex dependence of the downstream proton temperature anisotropy on the upstream alpha particle content and the shock inclination angle \citep{preisser_influence_2020}. Temperature anisotropy downstream of the shock can provide a source of free energy for the growth of mirror mode and Alfvén ion cyclotron mode waves \citep{mckean_magnetosheath_1996, kropotina_relaxation_2016, gary_ion_1993,gary_proton_1994}. These waves are deemed to cause further heating by scattering the directly transmitted ions. Upstream waves, if present, can also preheat the ions ahead of the shock ramp through interacting with the incident plasma \citep{scholer_whistler_2007}. In extreme cases, quasi-periodic plasma structures develop upstream of the shock that can contain heated plasma \citep{schwartz_quasi-parallel_1991,madanian_dynamics_2021,chen_solitary_2022}.

The ion heating and shock dynamics processes across low Mach number and very low $\beta$ quasi-perpendicular shocks is unique, as the magnetic energy density in the background plasma is already high and waves and perturbations upstream or downstream of the shock are relatively small amplitude. In this study, we investigate the dynamics of protons and heavy ions across the Earth's bow shock and analyze the bow shock response. The event is during the passage of the magnetic cloud which caused significant geomagnetic perturbations cause of which are not well understood \citep{zou_extreme_2024, despirak_geomagnetically_2024}. Therefore it demands careful analysis of different aspects of the interaction. We discuss the data sources in Section 2 and provide details of the observations in Section 3. In Section 4, we show the simulations of the ion dynamics. Discussion and Conclusions are provided in Sections 5 and 6, respectively.

\section{Data and Method}
We use measurements by the Magnetospheric Multiscale (MMS) spacecraft \citep{burch_magnetospheric_2016} of the Earth's bow shock. Data from the fast plasma investigation (FPI) \citep{pollock_fast_2016}, hot plasma composition analyzer (HPCA) \citep{young_hot_2016}, and the magnetometers \citep{russell_mms_2022} are utilized. More details of our processing of HPCA data are described in Appendix A. We also use measurements of the solar wind monitor Wind spacecraft far upstream of Earth to determine relevant upstream conditions \citep{lepping_wind_1995, lin_three-dimensional_1995, wilson_quarter_2021}. The clear signatures of the density peaks or enhancements in both Wind and MMS spacecraft enable determination of a proper time lag. The upstream He\textsuperscript{+} densities are derived indirectly from FPI measurements as described in Appendix B. Vector quantities, unless otherwise noted, are in the geocentric solar ecliptic (GSE) coordinate system in which $+x$ points towards the sun, $+y$ points opposite to the planetary motions in the ecliptic plane, and $+z$ is perpendicular to the ecliptic plane.

\section{Observations}
On 22 April 2023 a CME was recorded by the solar observatories near the Sun. The interplanetary shock, sheath, and the magnetic cloud associated with this event were observed $\sim40$ hours later by the solar wind monitor Wind spacecraft upstream of Earth (see Appendix D). The strong interplanetary magnetic field within the magnetic cloud results in an extremely low $\beta \sim 0.02$ and low Mach number $M_A \sim 1.9$ plasma. The solar wind flow eventually turned sub-Alfvénic and caused the formation of Alfvén wings around Earth \citep{chen_earths_2024}. We focus on the interaction of the magnetic cloud of the ICME with the Earth's bow shock hours after the ICME shock is observed. Motivated by this event and using a test particle model, \citet{graham_ion_2024} simulated the motion of different ion species across a perpendicular shock with a fixed jump ratio of $\sim 3.3$. They obtained downstream ion velocity distributions that resembled the observations reasonably well and noted the difficulties in characterizing different ion species in electrostatic analyzer data.

\begin{figure}
\centering
\includegraphics[width=\textwidth]{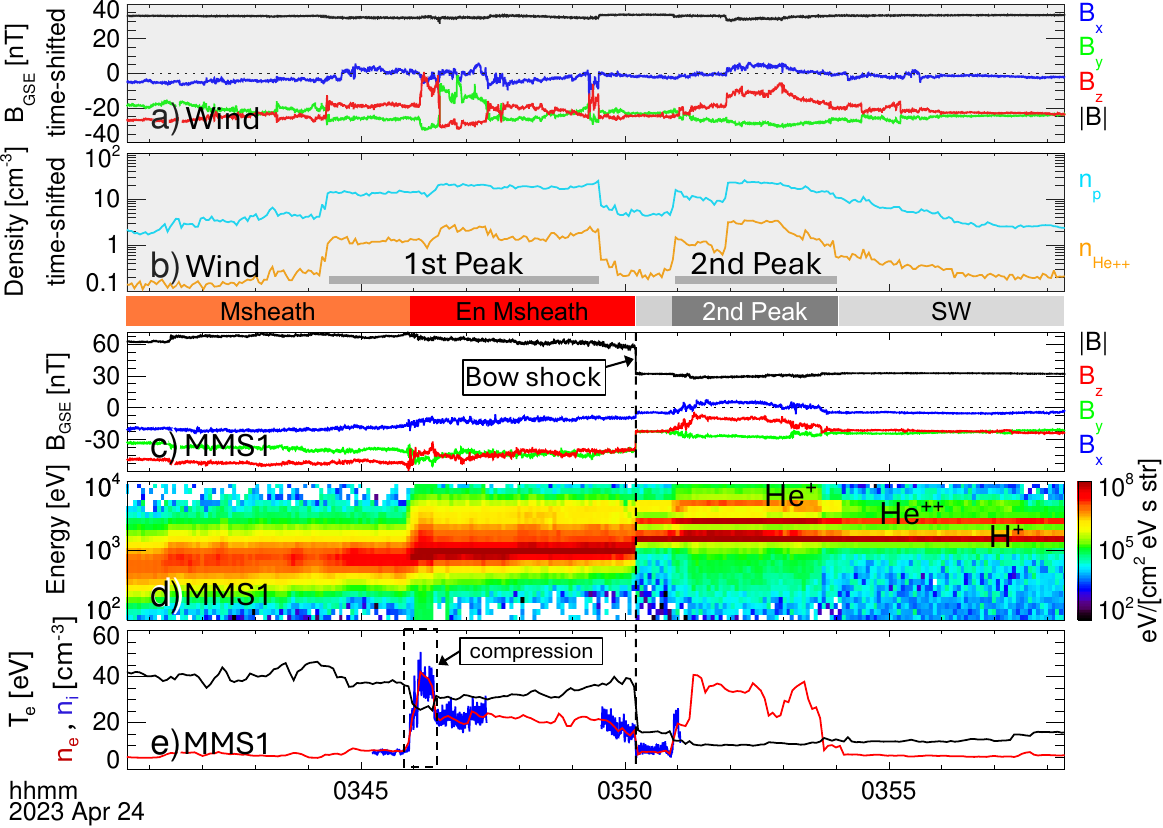}
\caption{An overview of the upstream density structure and its interaction with the bow shock on 24 April, 2023. Panels (a) and (b) show the magnetic field and  proton and alpha particle densities in the pristine solar wind. Magnetic field measurements (c), ion energy flux spectra (d), electron density (red), temperature (black), and ion density (blue) (e) across the bow shock are also shown. Density peaks of the double peak structure are annotated on panel (b). Data in panels (a) and (b) are from the Wind spacecraft time shifted by $\sim$ 42 minutes, while panels (c — e) show MMS1 spacecraft measurements. The regular magnetosheath (``Msheath"), enhanced magnetosheath ``En Msheath"), and solar wind (SW) regions in MMS data are identified above panel (c). The bow shock is marked with a dashed vertical line. The instance of initial plasma compression is identified by the dashed box on panel (e).}
\label{fig:fig1_mainoview}
\end{figure}

The magnetic cloud carries a “double-peak” density structure in which increased abundances of cold protons, alpha particles, and singly charged helium ions He\textsuperscript{+} are observed. Panels (a) and (b) in Figure \ref{fig:fig1_mainoview} (grey background) show measurements by the Wind spacecraft of the interplanetary magnetic field and solar wind plasma density. Several rotational discontinuities are evident in the magnetic field data within the first density peak where the proton and alpha particle densities reach 20.0 and 2.5 cm\textsuperscript{-3}, respectively (Figure \ref{fig:fig1_mainoview}b). Data in these panels are time-shifted to match MMS observations near the Earth's bow shock and to help identify corresponding structures downstream of the bow shock. The time lag ($\sim 42$ minutes) is determined by matching the onset of the second density peak. The solar wind plasma may evolve during the transport from L1 to the nose of the bow shock; However, signatures of such a large scale density structure (in duration and magnitude) are clearly visible in both MMS and Wind data. Enhanced abundances of helium group ions within the first density peak reduce the Alfvén speed and the bow shock Mach number increases to $M_A = 4.3$ which puts the shock in the supercritical regime and some proton reflection from the shock front is expected. More details about the upstream plasma parameters within the first density peak and the bow shock are listed in Table \ref{tab:tabl1}. The magnetic field vector is selected in the pristine solar wind and before the density structure onset (i.e., beginning of the interval in Figure \ref{fig:fig1_mainoview}a).

\begin{table}
 \caption{Upstream plasma and shock parameters during the first density peak in Figure \ref{fig:fig1_mainoview}.}
 \centering
 \begin{tabular}{l c c}
 \hline
  Parameter  & Value & Source  \\
 \hline
   $T_e, T_{H^+}, T_{He^{++}}, T_{He^+}$ [eV]  & 8.0, 8.5, 103, 103 &  MMS, Wind, Wind, Wind \\
   $n_e, n_{H^+}, n_{He^{++}}, n_{He^+}$ [cm\textsuperscript{-3}] & 25.6, 20, 2.5, 0.6 & - , Wind, Wind, MMS  \\
   $\textbf{V}_{H^+}$ [kms\textsuperscript{-1}] & (-580, 30, -60) & Wind   \\
   $\textbf{V}_{He^{++}}$ [kms\textsuperscript{-1}] & (-610, 30, -80) & Wind   \\
   $\textbf{V}_{He^{+}}$ [kms\textsuperscript{-1}] & (-610, 30, -80) & Wind   \\
   $\textbf{B}_{up}$ [nT] & (-4.47, -19.51, -26.35) &  Wind  \\
   $\theta_{Bn}$ [deg.]  & 68.5  & - \\
   $|\textbf{V}_{shock}|$ [kms\textsuperscript{-1}] & -129 & Appnd. C  \\
 \hline
 \end{tabular}
 \label{tab:tabl1}
 \end{table}

MMS measurements of the magnetic field and plasma in panels (c - e) of Figure \ref{fig:fig1_mainoview} begin inside the magnetosheath (``Msheath", orange segment). Between 03:46:05 and 03:50:12 UT when the spacecraft crosses the bow shock into the solar wind, both the ion energy spectrogram (panel (d)) and the total plasma density (the red line in panel (e)) show enhanced particle fluxes in the magnetosheath (``En Msheath", red segment). At the beginning of the ``En Msheath" period, the magnetic field strength is around 65 nT while the total plasma density increases to 45 cm\textsuperscript{-3}. These represent jumps by a factor of $\sim 2$ from the corresponding values within the upstream density peak. After the initial plasma compression (marked with a dashed rectangle on panel (e) of Figure \ref{fig:fig1_mainoview},) the plasma density decreases to around 20 cm\textsuperscript{-3} which is comparable to proton densities observed within the first density peak in the solar wind (``1st Peak" in panel (b)). The magnetosheath magnetic field strength (the black line in panel (c)) remains roughly constant during the plasma compression but it gradually decreases till the bow shock crossing. The plasma density variations within the ``En Msheath" region seem inconsistent with variations in the magnetic field strength and in contrast to the typical behavior of a fast mode compressional magnetohydrodynamic (MHD) shock wave \citep{lehmann_signatures_2016, delmont_parameter_2011}.

\begin{figure}
\centering
\includegraphics[width=0.75\textwidth]{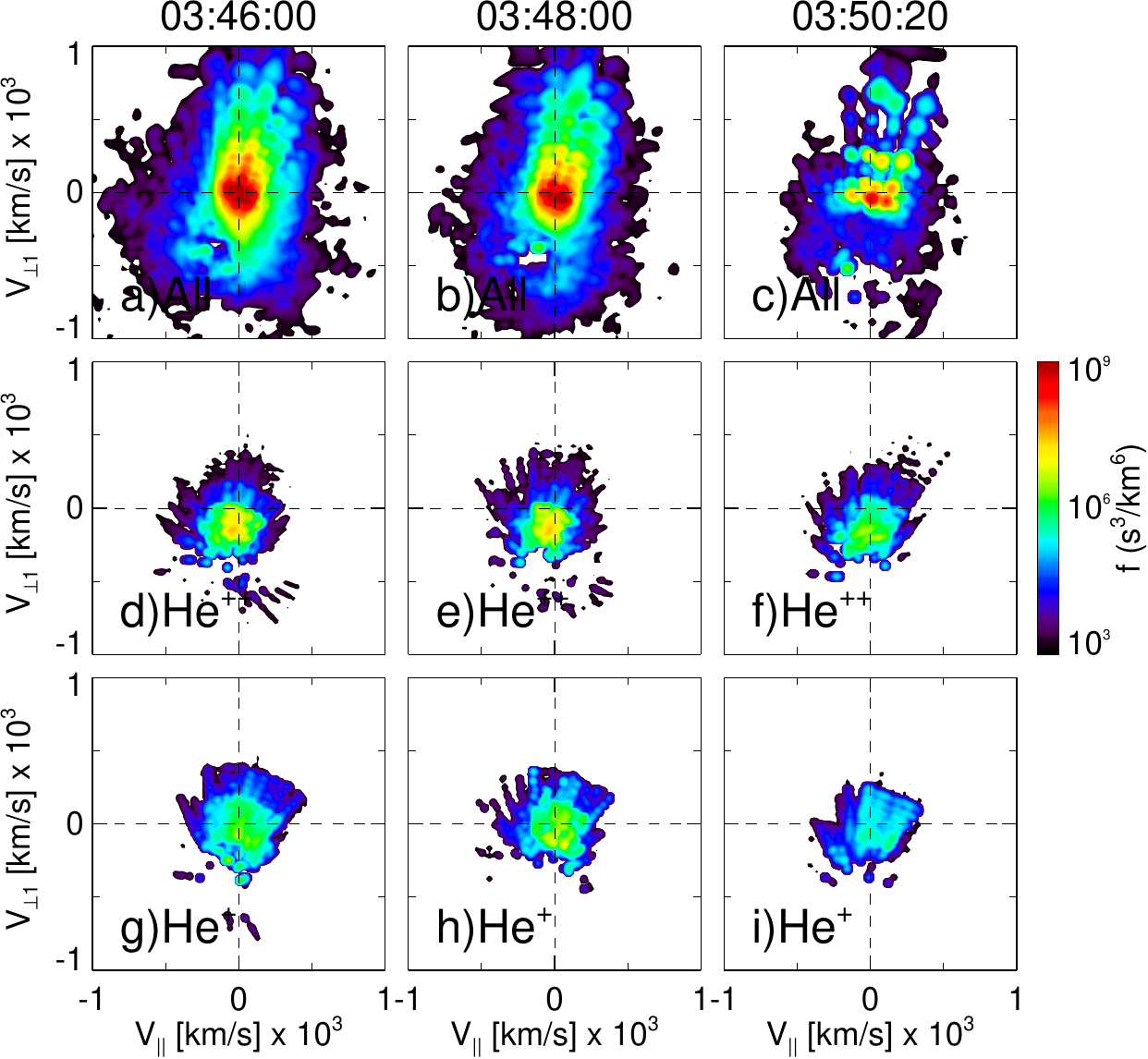}
\caption{2D phase space ion distributions in $V_{||} - V_{\perp1}$ plane downstream (first and second columns) and upstream (third column) of the bow shock. Data in the first row are from the FPI instrument while the second and third rows show data from the HPCA instrument. Distributions in panels (a-c) include all ions. Panels (d-f) show alpha particles, and panels (g-i) show He\textsuperscript{+} distributions. $V_{\perp1}$ is along the component of the population flow velocity perpendicular to the local magnetic field. }
\label{fig:fig2_3x3}
\end{figure}

At quasi-perpendicular shocks, heating and scattering of the solar wind proton beam mostly occur within the shock ramp layer \citep{thomsen_ion_1985, wilson_observations_2012,agapitov_energy_2023}, where electrostatic and electromagnetic waves decelerate the incident protons and scatter the ion beam in directions perpendicular to the magnetic field leading to highly anisotropic distributions in the magnetosheath. Figure \ref{fig:fig2_3x3} shows ion distributions in the $V_{||}-V_{\perp1}$ plane at three timestamps. Distributions at 03:46:00 and 03:48:00 UT are measured in the magnetosheath. The distributions in the third column show the solar wind beam upstream of the bow shock for comparison. FPI measurements in the first row include all ions in the plasma regardless of the ion mass or charge state. Thus the distributions in panels (a - b) of Figure \ref{fig:fig2_3x3} show a heated solar wind plasma and consist of protons laced with fluxes of alpha particles and He\textsuperscript{+} ions. Protons are still much more abundant compared to heavier ions even within the density peak. Signs of heating and scattering of the core proton beam is evident in the distributions in panels (a) and (b) caused by electrostatic and electromagnetic wave fields at the shock front \citep{agapitov_energy_2023, wilson_observations_2012}, as well as dispersion by the cross shock electrostatic potential. The most noticeable difference between the distribution in panel (a) when the initial density compression is observed and the distribution in panel (b) is in the intensity of the core population which is higher during the compression period.

Alpha particles and He\textsuperscript{+} ions are distinguished in HPCA measurements in the second and third rows. The core population of these ions also shows higher fluxes in the magnetosheath compared to the upstream distribution. However, for alpha particles and He\textsuperscript{+} ions, scattering around the core population is much less compared to protons. This is expected as the kinetic scale of heavy ions is much larger than the shock thickness ($\sim$ a proton convective gyroradius) and larger than the spatial scales of electromagnetic waves within the shock layer. The interaction of upstream charged particles with the shock is dependent on the particle’s charge state and mass. Waves can accelerate electrons on much smaller scales than ions \citep{chen_electron_2018}, but ions carry more of the energy than electrons \citep{lei_quantified_2024,schwartz_energy_2022}. While heavy ions are decelerated along the shock normal by the cross shock electrostatic potential, they are less scattered by wave fields compared to protons.

\begin{figure}
\centering
\includegraphics[width=0.75\textwidth]{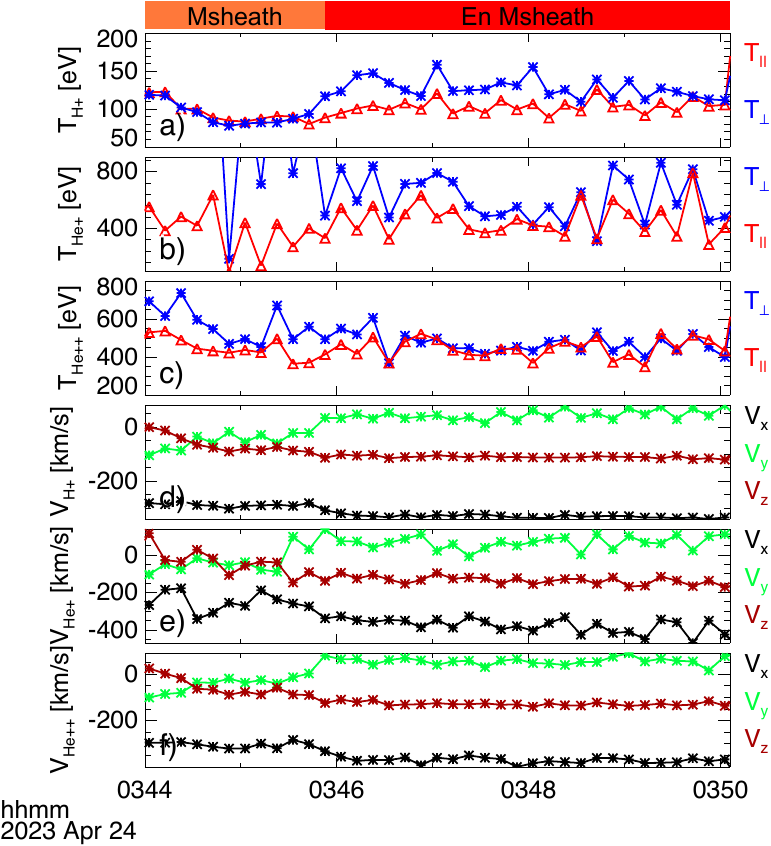}
\caption{(a - c) Downstream temperatures for protons, He\textsuperscript{+}, and alpha particles, respectively. Parallel (red) and perpendicular (blue) temperatures are shown. (d - e) Downstream ion velocities for protons, He\textsuperscript{+}, and alpha particles, respectively.}
\label{fig:fig3_hpcats}
\end{figure}

Figure \ref{fig:fig3_hpcats} shows ion temperatures and velocities downstream of the bow shock. Protons within the ``En Msheath" period exhibit perpendicular temperature anisotropy as is evident in panel (a). He\textsuperscript{+} ions and alpha particles in panels (b) and (c) also show anisotropic distributions with instances of high perpendicular temperatures near the beginning and end of the interval and less anisotropy in the middle. The HPCA estimates of ion temperature are sensitive to the ion abundance and the measured count rates. As such, before the ``En Msheath" period (i.e., before the first density peak with high abundances of heavy ions passes through the bow shock,) He\textsuperscript{+} temperatures are noisy. At $\sim$03:46:10 UT when the initial plasma density compression is observed (Figure \ref{fig:fig1_mainoview}e), the average ion temperature for protons, He\textsuperscript{+}, and alpha particles is 124, 605, and 487 eV, respectively. Each ion species also travels at somewhat different velocities downstream of the bow shock as shown in panels (d-f) of Figure \ref{fig:fig3_hpcats}. There is a clear increase in the anti-sunward flow velocity ($|V_x|$) near the ``En Msheath" period. This feature appears in the He\textsuperscript{+} velocities first, followed by a similar increase in alpha particle velocities and then in proton velocities.

\section{Hybrid Simulations}
We perform a series of hybrid simulations to study the multi-ion dynamics downstream of the low $\beta$ shock event in Figure \ref{fig:fig1_mainoview}. We use 1D Hybrid-VPIC which treats ions kinetically and electrons as a massless fluid. The initial condition consists of uniform plasma and electromagnetic fields, with $B_x = B_{up} \cos \theta_{Bn}$, $B_y = 0$, $B_z = B_{up} \sin \theta_{Bn}$, and $E_y = -V B_{up} \sin \theta_{Bn}$. The initial plasma moves in the negative $x$ direction with velocity $-V$. The lower $x$ boundary uses conducting walls and reflects particles, while plasma is injected at the upper $x$ boundary. The simulation domain is 200 $d_i =$ 2000 cells, and is initialized with 200 particles per species per cell. The physical parameters follow Table \ref{tab:tabl1}, transformed into the downstream rest frame. We note that the ion temperatures throughout the density structure change from $\sim 50$ eV at the onset of the enhancement to 103 eV at the peak density point. With fixed initial conditions in the simulations, we choose a temperature of 70 eV for He\textsuperscript{+} and alpha particles to reflect the mean temperature of the whole structure. The injected ion velocity at the right boundary was 303 kms\textsuperscript{-1} for all species. The rightward moving shock front was 131 kms\textsuperscript{-1} from the simulation box, so that the inflow speed in the shock rest frame is 434 kms\textsuperscript{-1}.

\begin{figure}[h]
\centering
\includegraphics[width=0.75\textwidth]{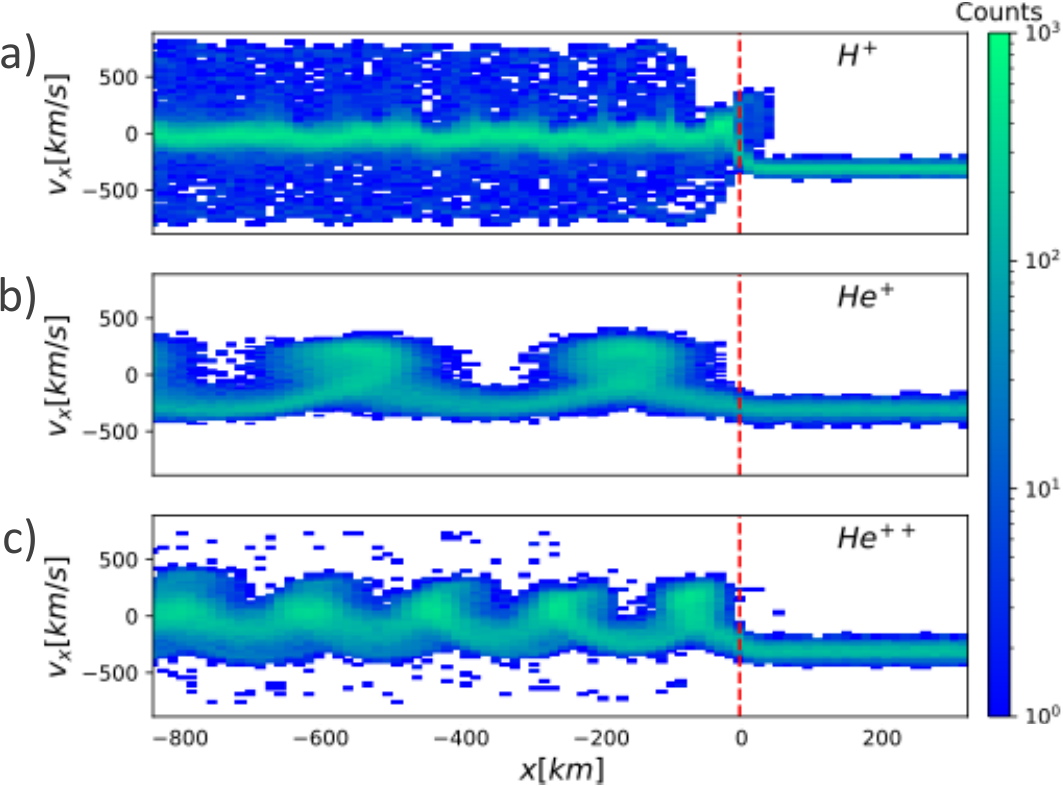}
\caption{Counts of simulated particles for a) protons, b) He\textsuperscript{+}, and c) alpha particles.}
\label{fig:fig4_sims}
\end{figure}

The results after $t \Omega_{cH} = 40$ are shown in Figure \ref{fig:fig4_sims}, where the ion $x$-$v_x$ phase space distributions are shown. The shock front is at $x = 0$ marked by the dashed line. Only protons show a significant population of reflected ions (blue colors in Figure \ref{fig:fig4_sims}a). Upon crossing the shock and due to a relative change in their rest frame velocity, He\textsuperscript{+} ions and alpha particles begin to gyrate in the enhanced downstream magnetic field and exhibit quasi periodic gyrophase bunching. The ion temperatures during the bunching phase will appear higher than other times.

\section{Discussion}
In the context of magnetohydrodynamic (MHD) shock waves, three classes of linear wave modes, namely fast ($v_f$), intermediate ($v_i$), and slow ($v_s$), can be obtained from the small perturbation dispersion relation. The intermediate (co-planar) shock is purely a transverse mode, while fast and slow modes are compressive waves. These waves travel at phase speeds of: 

\begin{eqnarray}
    v_f = \sqrt{\frac{1}{2} (v_{ms}^2 + \sqrt{v_{ms}^4 - 4 c_s^2 v_{Alf}^2cos^2\theta_{Bn}})} \\
    v_i = v_{Alf} cos\theta_{Bn} \\
    v_s = \sqrt{\frac{1}{2} (v_{ms}^2 - \sqrt{v_{ms}^4 - 4 c_s^2 v_{Alf}^2cos^2\theta_{Bn}})}
\end{eqnarray}

\noindent where $v_{ms} = \sqrt{v_{Alf}^2 + c_s^2}$ is the magnetosonic wave speed, $v_{Alf} = |\textbf{B}| / \sqrt{\mu_{0}\rho}$ is the Alfvén speed, and $c_s = \sqrt{\gamma P / \rho}$ is the sonic wave speed in the plasma. In these equations, $\gamma$ is the adiabatic index, $\mu_{0}$ is the vacuum permeability, $\theta_{Bn}$ is measured using local magnetic field vectors. The total mass density $\rho$ is defined as $\rho = m_{H^+}[n_{H^+} + 4(n_{He^+}+n_{He^{++}})]$, and $P$ includes the thermal pressure of all ions and electrons. 

In Table \ref{tab:tabl2} we examine the characteristic wave speeds to ion flow velocities along the shock normal ($v_n$) and perpendicular to the magnetic field ($v_{\perp}$) for three cases: 1) during the ``En Msheath" period with the initial plasma compression, 2) during the ``En Msheath" period for non-compressed plasma, 3) within the density peak upstream of the bow shock. For case (1) we obtain the downstream ion densities from the conservation of particle flux across the shock (i.e., $n_{H^+}=26.7$, $n_{He^{++}}=3.1$, $n_{He^+}=0.8$). Ion densities and fractional abundances of three ion species for cases (2) and (3) are listed in Table \ref{tab:tabl1}. As discussed below Figure \ref{fig:fig1_mainoview}, after the initial compression, the magnetosheath plasma densities are comparable to densities within the first peak in the solar wind. In cases (1) and (2), local downstream flow velocity for protons is $\textbf{V}_{H^+}=(-326,40,-106)$ kms\textsuperscript{-1}, and $\textbf{V}_{He^+}=(-345,63,-114)$ kms\textsuperscript{-1} for He\textsuperscript{+}, and $\textbf{V}_{He^{++}}=(-367,57,-119)$ kms\textsuperscript{-1} for alpha particles. The interaction of ion species with different mass per charge ratios with the electromagnetic and electrostatic fields at the shock front is a kinetic process (i.e., on scales of ion gyroradius). However, for simplicity and for the lack of a better theory that accounts for such kinetic effects, in this discussion we focus on the fluid properties of ion populations.

\begin{table}[hb]
 \caption{Comparison of characteristic wave speeds and plasma flow speeds. All entries are in units of kms\textsuperscript{-1}.}
 \centering
 \begin{tabular}{c c c c}
 \hline
     & \multicolumn{2}{c}{En Msheath} & Upstream \\
 \hline
     & 1: Compression  & 2: non-Compression  & 3: Peak \\
 \hline
   $v_{Alf.}$ & 219 & 249 & 127 \\
   $c_{s}$ & 148 & 146 & 54 \\
   $v_f$  & 260 &  286 & 137 \\
   $v_s$ & 44 & 45 & 18 \\
   $v_i$ & 78 & 89 & 46 \\
   $v_{n} (H^+, He^+, He^{++})$ & \multicolumn{2}{c}{267, 291, 305} & 357, 377, $377^{*}$ \\
   $v_{\perp}(H^+, He^+, He^{++})$ &  \multicolumn{2}{c}{330, 357, 375} & - \\
 \hline
 \multicolumn{4}{l}{$^*$ The upstream velocities are in the normal incidence frame (NIF) (Appendix C).}
\end{tabular}
\label{tab:tabl2}
\end{table} 

The characteristic wave speeds and flow velocities in Table \ref{tab:tabl2} indicate that within the ``En Msheath" period and during the initial plasma compression, the flow of ion species along the shock normal downstream of the bow shock remains super Alfvénic. Additionally, ion motions perpendicular to the magnetic field are still super fast magnetosonic (i.e. speeds higher than $v_f$). After the initial compression, the proton flow velocity along the shock normal is sub fast. However, ions, particularly He\textsuperscript{+} and alpha particles, still travel at super magnetosonic speeds perpendicular to the magnetic field (case 2). Increased ion fluxes within the ``En Msheath" plasma reduce the characteristic waves speeds in that region as compared to the regular magnetosheath period. Thus, we observe a transition state in which the bow shock appears as an irregular nonstationary boundary where the plasma flow is super Alfvénic both upstream and downstream of the bow shock. 

As we discussed in Figure \ref{fig:fig1_mainoview}, the plasma density compression downstream of the bow shock is either decreased, suppressed, or is inconsistent with the magnetic field jump across the shock. To our knowledge, this is the first report of such a shock structure and compression pattern. Heavy ions (e.g., He\textsuperscript{+} and alpha particles) in the upstream solar wind carry a significant portion of the upstream flow energy ($\sim 40$ \%). From the energy conservation across a low Mach number perpendicular shock, we can write for the downstream speed of an ion species $s$ with mass $m_s$ and charge $q_s$ as $v_s = V_{s_{up,NIF}} \sqrt{1-(\epsilon_s q_s/m_s )}$, where $\epsilon_s$ is the cross-shock electrostatic potential potential normalized by the upstream flow energy. Using each species' upstream an downstream velocities listed in Tables \ref{tab:tabl1} and \ref{tab:tabl2}, we find normalized cross-shock potentials of $\epsilon_{H^+} = 0.44$, $\epsilon_{He^+} = 0.40$, and $\epsilon_{He^{++}} = 0.34$. Comparable $\epsilon$ values for different species indicate that all ions are decelerated consistently across the shock. As such, the shock does not seem to repartition the upstream energy (e.g., heating the protons only) and each ion species seemingly carries a similar fraction of their upstream energy in the downstream in the form of either thermal or ram energy. Even during the initial plasma compression period, the associated change in the proton thermal pressure is only a fraction of the magnetic energy density and does not cause significant perturbations in the background magnetic field. Additionally, as the simulations in Figure \ref{fig:fig4_sims} show, after crossing the shock, heavy ions gyrophase bunch at different distances to the shock and do not contribute to the downstream thermal energy concurrent to protons.

The distributions in Figure \ref{fig:fig2_3x3} show that heavy ions are not efficiently scattered at the shock. We obtain downstream to upstream temperature ratios of 14.6, 5.8, and 4.7 for protons, He\textsuperscript{+} ions, and alpha particles, respectively. The downstream to upstream temperature ratio, a crude proxy of how much heating each ion species experienced, is much lower for He\textsuperscript{+} and alpha particles compared to that of the protons. Temperature ratios do not seem to follow a mass per charge dependent relation, but they could be considered mass dependent. We associate the much higher change in the temperature of protons to ion reflection and scattering by electromagnetic an electrostatic wave fields at the shock front. Both of these processes are absent for heavy ions, which require high amplitude electromagnetic perturbations to be reflected or scattered \citep{madanian_direct_2021,caprioli_chemical_2017, broll_mms_2018}. As such, the dissipation of heavy ion kinetic energy is through downstream processes affecting the directly transmitted ions rather than at the shock front \citep{fuselier_h_1994}. Plasma compression by a decelerating potential, adiabatic heating, and gyrophase bunching can also increase the ion temperature. However, these reversible processes do not increase the plasma entropy. Since heavy ions in the solar wind are in general hotter than protons, upon crossing the bow shock, they will also possess higher temperatures. One thus can argue that heavy ions carry significant portion of the downstream enthalpy. As at the heliospheric termination shock, hot pickup ions account for a larger fraction of thermal energy flux downstream of the shock despite their lower densities \citep{wu_energy_2009}. 

The presence of heavy ions in the upstream plasma poses a challenge to both observations and numerical simulations of shocks. A proper determination of the relative amplitude of the wave fields to the static electric field in simulations is important to avoid overestimation of the heavy ion heating rates across the shock \citep{krasnoselskikh_dynamic_2013, wilson_discrepancy_2021}.

\section{Conclusion}

In this study we investigate the dynamics of protons, alpha particles and singly charged helium ions observed in the prominence plasma carried by the magnetic cloud of an ICME with the Earth's bow shock. The energy balance across the shock is strongly perturbed in the presence of the prominence plasma. The strong magnetic field within the flux rope of the magnetic cloud results in a low Mach number and very low $\beta$ solar wind plasma where magnetic perturbations due to particle dynamics have relatively low amplitudes. The bow shock appears inefficient in heating the solar wind and weakly shocked heavy ions and protons continue to flow at super Alfvénic speeds in the downstream and within the sheath plasma. The density structure discussed here caused a noticeable space weather impact on Earth producing significant geomagnetically induced currents \citep{zou_extreme_2024,despirak_geomagnetically_2024}. Analysis of the structure impact on the magnetosheath, magnetopause, and their links to the space weather effects is left for future investigations. 

Through a multi-component plasma and a planar shock hybrid simulation, we observe differential kinetic behaviors of various components over a large downstream region. The simulation results are from a 1D hybrid run with fixed upstream conditions. The results intend to reveal the different dynamics of heavy ions compared to protons downstream of a shock. This is a preliminary attempt to approach this problem and detailed multidimensional full kinetic and hybrid simulations are needed in future studies to illustrate the effects of heavy ions on the shock layer in a proton-dominant plasma.


\begin{acknowledgments}
This research was supported in part by the MMS project, and by the NASA MMS Early Career grant 80NSSC23K0009.
\end{acknowledgments}

\noindent Data and Code Availability 
Data products used in this study are available through public archives at MMS Science Data Center at \url{https://lasp.colorado.edu/mms/sdc/public/}. The following data products are used: FPI burst and fast mode moments and distributions \citep{gershman_mms_2022,gershman_mms_2022-1,gershman_mms_2022-2}, HPCA survey distributions \citep{fuselier_mms_2022}, electric field  \citep{ergun_mms_2022}, magnetic field \citep{russell_mms_2022,russell_mms_2022-1}, and Wind-3DP PM \citep{koval_wind_2021}.

%

\appendix
\section{HPCA Data Calibration}

The HPCA has a measurement cycle of 10 s and the full sky coverage is obtained with the spacecraft spin. We accurately reprocess and recalibrate HPCA data to reduce the effects of noise and uncertainties in the data. HPCA measurements can be sensitive to the plasma flux direction resulting in a high degree of variability due to ion flux energies not being properly resolved by the instrument (e.g., the cold solar wind ion distribution). Our usage of HPCA data is limited to the magnetosheath plasma measurements.
\subsection{Data Contamination}
When HPCA measures environments with high proton fluxes, such as the magnetosheath and solar wind, two types of contamination affects the HPCA distributions: 1) Bleed over protons, which refer to uncorrelated coincidence counts recorded in the TOF section of the instrument at energies corresponding to the bulk proton energy, and 2) Deflected ions, which refer to ions that deflect off of the inner ESA or grid and are recorded as counts corresponding to the opposite direction of the incident ions. 
\subsection{Bleed Over Proton Removal}
Bleed over protons affect both He\textsuperscript{+} and He\textsuperscript{++} distributions, however since the bulk H\textsuperscript{+} energy is close to the bulk He++ energy, we are unable to separate bleed over counts from real counts in the He\textsuperscript{++} distributions. On the other hand, for the He\textsuperscript{+} distributions we remove bleed over proton counts by setting the phase space density to 0 for azimuth angles within $30^{\circ}$ of the bulk flow direction and within a defined energy range which depends on the bulk energy of the protons (e.g., 100 – 1000 eV in the solar wind). Note that the bulk energy of He\textsuperscript{+} is four times that of the H\textsuperscript{+} and so this process effectively removes bleed over counts, while minimally affecting real He\textsuperscript{+} counts. Further details on bleed over proton contamination in HPCA ion distributions are discussed in \citep{gomez_extramagnetospheric_2019, starkey_mms_2020,starkey_acceleration_2019}.

\subsection{Deflected Ion Removal}
These ions result in counts corresponding to the opposite direction of travel of the incident ion population. For example, in the solar wind these deflected ions are recorded in the anti-sunward direction over a wide range of energies and across all elevations (due to the internal scattering). These counts are removed by setting the phase space density to 0 for azimuths within 60$^{\circ}$ of the anti-solar wind direction, all elevations, and all energies.

\section{FPI Partial Moments}

In this section we discuss our method of determining the upstream parameters for He\textsuperscript{+} ions listed in Table \ref{tab:tabl1}. In the absence of independent He\textsuperscript{+} measurements in the solar wind, we use FPI data in the solar wind to calculate partial moments of the ion distributions and determine the density of He\textsuperscript{+} ions. The FPI instrument is a set of electrostatic analyzers measuring charged particles without differentiating the particle's mass. The measured energies are indeed the energy per charge of the particles. As such, both alpha particles and singly charged helium ions appear as additional peaks in the electrostatic analyzer measurements of the solar wind. These are labeled as ``He\textsuperscript{++}" and ``He\textsuperscript{+}" in FPI ion energy spectra in panel (d) of Figure \ref{fig:fig1_mainoview} between 03:51:00 and 03:53:40 UT corresponding to the second density peak in the solar wind. This period is used for the partial moment analysis.

The partial density moment is calculated by simply integrating the energy flux distribution ($\psi$) over the specific energy range occupied by each ion species $s$: 
\begin{equation}
    n_{s} = \sqrt{\frac{m_{H^+}}{2}} \int\int^{E_{max}}_{E_{min.}}{\frac{\psi(E,\omega)}{E^{1.5}} dE d \omega}
\end{equation}

\noindent We then account for different mass ($M_s/m_{H^+}=\mu_s$) and charge ($q_s/e=g_s$) of the ion $s$ with respect to protons by propagating the corresponding coefficients throughout the integration. Throughout the text, we compare plasma conditions between upstream and downstream for the first density peak which is observed by MMS downstream of the bow shock. Since the density for protons and alpha particles appear to be comparable between the two peaks in the solar wind (Figure \ref{fig:fig1_mainoview}b), it is reasonable to expect that a similar abundance of He\textsuperscript{+} ions exists in the first density peak. It should be noted that FPI is not designed to measure the solar wind plasma and higher order moments (e.g., temperature and pressure) even for protons become less reliable or not reliable at all. We assume He\textsuperscript{+} ions in the solar wind have the same temperature and velocity as alpha particles, which we obtained from the Wind spacecraft measurements.


\section{Bow Shock Properties and Frame Definitions}
We determine the bow shock properties when MMS spacecraft observe the ``En Msheath" period. The shock normal vector ($\hat{n}$) is determined from the mixed coplanarity method \citep{schwartz_analysis_1998} using upstream and downstream proton flow velocity and magnetic field vectors selected before the onset of the density peaks and the ``En Msheath" period, respectively (see Table \ref{tab:tabl1} and Figure \ref{fig:fig1_mainoview}). This is to avoid multiple magnetic field rotations within the density peak. We also note that the bow shock geometry at the position of MMS remains quasi-perpendicular for all IMF vectors shown in Figure \ref{fig:fig1_mainoview}a. Significantly different shock geometries lead to asymmetric passage of upstream structures through the bow shock \citep{madanian_asymmetric_2022}. We obtain two estimates for the speed of the receding bow shock. A bow shock speed of 109 kms\textsuperscript{-1} is obtained by timing analysis of the four MMS spacecraft measurements of the bow shock crossing at 03:50:11 UT which could be different than when the first density peak initially hit the bow shock. The timing analysis between MMS and THEMIS-D spacecraft, positioned several Earth radii downstream and also observing the receding bow shock, results in a bow shock speed of 150 kms\textsuperscript{-1}. We take the average of these two estimates as the bow shock speed of $V_{shock}=$129 kms\textsuperscript{-1}. To determine the incident flow velocity along the shock normal (Table 2), we transform the ion velocity vectors to the normal incidence frame (NIF) through $\textbf{V}_{up,NIF} = \textbf{V}_{up} - \textbf{V}_{NIF}$ where $\textbf{V}_{NIF}=\hat{n} \times (\hat{n} \times \textbf{V}_{up})$ and $\textbf{V}_{up} = \textbf{V}_{up,sc} - \textbf{V}_{shock}$. The NIF moves parallel to the shock surface. With proper transformation, the normal component of an ion velocity vector downstream of the shock in the NIF is the same as that in the spacecraft frame.

\section{Supplementary Figure}
\begin{figure}
\centering
\includegraphics[width=0.7\textwidth]{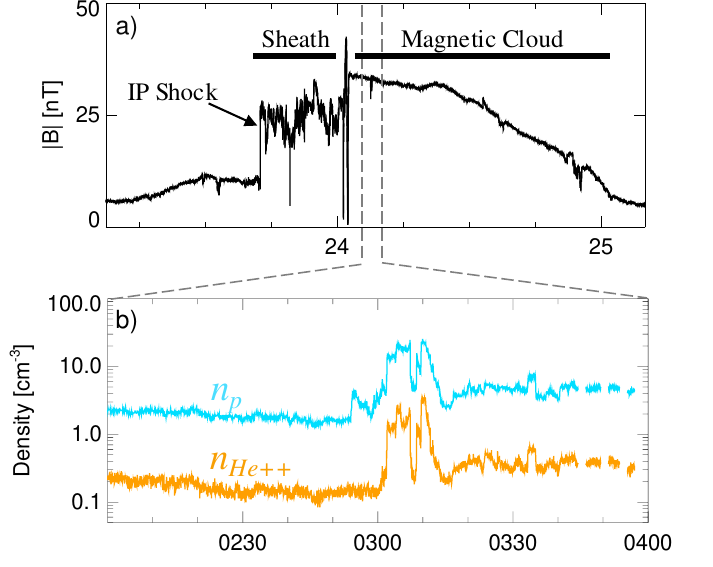}
\caption{Wind spacecraft observations of the ICME. Panel (a) shows magnetic field strength measurements over two days starting on 23 April 2023. Different components of the ICME, namely the interplanetary (IP) shock, sheath, and the magnetic cloud are annotated on this panel. Proton (cyan) and alpha particle (yellow) densities in panel (b) are shown during the period of the double peak density structure within the magnetic cloud associated with the prominence plasma.}
\label{fig:supp_icme}
\end{figure}

Figure \ref{fig:supp_icme} shows the structure of the entire ICME event and the instance of the density structure observed by the solar wind monitor far upstream of Earth.

\bibliography{references}

\begin{thebibliography}{}
\expandafter\ifx\csname natexlab\endcsname\relax\def\natexlab#1{#1}\fi
\providecommand{\url}[1]{\href{#1}{#1}}
\providecommand{\dodoi}[1]{doi:~\href{http://doi.org/#1}{\nolinkurl{#1}}}
\providecommand{\doeprint}[1]{\href{http://ascl.net/#1}{\nolinkurl{http://ascl.net/#1}}}
\providecommand{\doarXiv}[1]{\href{https://arxiv.org/abs/#1}{\nolinkurl{https://arxiv.org/abs/#1}}}

\bibitem[{Agapitov {et~al.}(2023)Agapitov, Krasnoselskikh, Balikhin, Bonnell, Mozer, \& Avanov}]{agapitov_energy_2023}
Agapitov, O.~V., Krasnoselskikh, V., Balikhin, M., {et~al.} 2023, The Astrophysical Journal, 952, 154, \dodoi{10.3847/1538-4357/acdb7b}

\bibitem[{Bame {et~al.}(1968)Bame, Hundhausen, Asbridge, \& Strong}]{bame_solar_1968}
Bame, S.~J., Hundhausen, A.~J., Asbridge, J.~R., \& Strong, I.~B. 1968, Physical Review Letters, 20, 393, \dodoi{10.1103/PhysRevLett.20.393}

\bibitem[{Broll {et~al.}(2018)Broll, Fuselier, Trattner, Schwartz, Burch, Giles, \& Anderson}]{broll_mms_2018}
Broll, J.~M., Fuselier, S.~A., Trattner, K.~J., {et~al.} 2018, Geophysical Research Letters, 45, 49, \dodoi{10.1002/2017GL075411}

\bibitem[{Burch {et~al.}(2016)Burch, Moore, Torbert, \& Giles}]{burch_magnetospheric_2016}
Burch, J.~L., Moore, T.~E., Torbert, R.~B., \& Giles, B.~L. 2016, Space Science Reviews, 199, 5, \dodoi{10.1007/s11214-015-0164-9}

\bibitem[{Burgess {et~al.}(2016)Burgess, Hellinger, Gingell, \& Trávníček}]{burgess_microstructure_2016}
Burgess, D., Hellinger, P., Gingell, I., \& Trávníček, P.~M. 2016, Journal of Plasma Physics, 82, 905820401, \dodoi{10.1017/S0022377816000660}

\bibitem[{Burlaga {et~al.}(1998)Burlaga, Fitzenreiter, Lepping, Ogilvie, Szabo, Lazarus, Steinberg, Gloeckler, Howard, Michels, Farrugia, Lin, \& Larson}]{burlaga_magnetic_1998}
Burlaga, L., Fitzenreiter, R., Lepping, R., {et~al.} 1998, Journal of Geophysical Research: Space Physics, 103, 277, \dodoi{10.1029/97JA02768}

\bibitem[{Cane {et~al.}(1986)Cane, Kahler, \& Sheeley}]{cane_interplanetary_1986}
Cane, H.~V., Kahler, S.~W., \& Sheeley, N.~R. 1986, Journal of Geophysical Research: Space Physics, 91, 13321, \dodoi{10.1029/JA091iA12p13321}

\bibitem[{Caprioli {et~al.}(2017)Caprioli, Yi, \& Spitkovsky}]{caprioli_chemical_2017}
Caprioli, D., Yi, D.~T., \& Spitkovsky, A. 2017, Physical Review Letters, 119, 171101, \dodoi{10.1103/PhysRevLett.119.171101}

\bibitem[{Chen {et~al.}(2022)Chen, Halekas, Wang, DiBraccio, Romanelli, Ng, Russell, Schwartz, Sibeck, Farrell, Pollock, Gershman, Giles, \& Collado‐Vega}]{chen_solitary_2022}
Chen, L., Halekas, J., Wang, S., {et~al.} 2022, Geophysical Research Letters, \dodoi{10.1029/2021GL097600}

\bibitem[{Chen {et~al.}(2024)Chen, Gershman, Burkholder, Chen, Sarantos, Jian, Drake, Dong, Gurram, Shuster, Graham, Le~Contel, Schwartz, Fuselier, Madanian, Pollock, Liang, Argall, Denton, Rice, Beedle, Genestreti, Ardakani, Stanier, Le, Ng, Bessho, Pandya, Wilder, Gabrielse, Cohen, Wei, Russell, Ergun, Torbert, \& Burch}]{chen_earths_2024}
Chen, L., Gershman, D., Burkholder, B., {et~al.} 2024, Geophysical Research Letters, 51, e2024GL108894, \dodoi{10.1029/2024GL108894}

\bibitem[{Chen {et~al.}(2018)Chen, Wang, Wilson, Schwartz, Bessho, Moore, Gershman, Giles, Malaspina, Wilder, Ergun, Hesse, Lai, Russell, Strangeway, Torbert, F.-Vinas, Burch, Lee, Pollock, Dorelli, Paterson, Ahmadi, Goodrich, Lavraud, Le~Contel, Khotyaintsev, Lindqvist, Boardsen, Wei, Le, \& Avanov}]{chen_electron_2018}
Chen, L.-J., Wang, S., Wilson, L., {et~al.} 2018, Physical Review Letters, 120, \dodoi{10.1103/PhysRevLett.120.225101}

\bibitem[{Delmont \& Keppens(2011)}]{delmont_parameter_2011}
Delmont, P., \& Keppens, R. 2011, Journal of Plasma Physics, 77, 207, \dodoi{10.1017/S0022377810000115}

\bibitem[{Despirak {et~al.}(2024)Despirak, Setsko, Lubchich, Hajra, Sakharov, Lakhina, Selivanov, \& Tsurutani}]{despirak_geomagnetically_2024}
Despirak, I., Setsko, P., Lubchich, A., {et~al.} 2024, Journal of Atmospheric and Solar-Terrestrial Physics, 261, 106293, \dodoi{10.1016/j.jastp.2024.106293}

\bibitem[{Ellacott \& Wilkinson(2003)}]{ellacott_heating_2003}
Ellacott, S.~W., \& Wilkinson, W.~P. 2003, Journal of Geophysical Research: Space Physics, 108, 2003JA009919, \dodoi{10.1029/2003JA009919}

\bibitem[{Ergun {et~al.}(2022)Ergun, Lindqvist, Torbert, Ahmadi, Graham, \& Burch}]{ergun_mms_2022}
Ergun, R.~E., Lindqvist, P.-A., Torbert, R.~B., {et~al.} 2022, {MMS} 2 {Electric} {Double} {Probe} ({EDP}) {Three}-{Dimensional} {Electric} {Field}, {Level} 2 ({L2}), {Burst} {Mode}, 0.1220703125 ms {Data},  NASA Space Physics Data Facility, \dodoi{10.48322/PKTT-WW31}

\bibitem[{Fuselier \& Schmidt(1994)}]{fuselier_h_1994}
Fuselier, S.~A., \& Schmidt, W. K.~H. 1994, Journal of Geophysical Research, 99, 11539, \dodoi{10.1029/94JA00350}

\bibitem[{Fuselier {et~al.}(2022)Fuselier, Young, Gomez, \& Burch}]{fuselier_mms_2022}
Fuselier, S.~A., Young, D.~T., Gomez, R.~G., \& Burch, J.~L. 2022, {MMS} 1 {Hot} {Plasma} {Composition} {Analyzer} ({HPCA}) {Ions}, {Level} 2 ({L2}), {Survey} {Mode}, 0.625 s {Data},  NASA Space Physics Data Facility, \dodoi{10.48322/HJVZ-AW21}

\bibitem[{Gary {et~al.}(1993)Gary, Fuselier, \& Anderson}]{gary_ion_1993}
Gary, S.~P., Fuselier, S.~A., \& Anderson, B.~J. 1993, Journal of Geophysical Research: Space Physics, 98, 1481, \dodoi{10.1029/92JA01844}

\bibitem[{Gary {et~al.}(1994)Gary, McKean, Winske, Anderson, Denton, \& Fuselier}]{gary_proton_1994}
Gary, S.~P., McKean, M.~E., Winske, D., {et~al.} 1994, Journal of Geophysical Research: Space Physics, 99, 5903, \dodoi{10.1029/93JA03583}

\bibitem[{Gedalin(1997)}]{gedalin_ion_1997}
Gedalin, M. 1997, Surveys in Geophysics, 18, 541, \dodoi{10.1023/A:1006509702173}

\bibitem[{Gershman {et~al.}(2022{\natexlab{a}})Gershman, Giles, Pollock, Moore, Kreisler, \& Burch}]{gershman_mms_2022}
Gershman, D.~J., Giles, B.~L., Pollock, C.~J., {et~al.} 2022{\natexlab{a}}, {MMS} 1 {Fast} {Plasma} {Investigation}, {Dual} {Ion} {Spectrometer} ({FPI}, {DIS}) {Distribution} {Moments}, {Level} 2 ({L2}), {Burst} {Mode}, 0.15 s {Data},  NASA Space Physics Data Facility, \dodoi{10.48322/QGGF-VR83}

\bibitem[{Gershman {et~al.}(2022{\natexlab{b}})Gershman, Giles, Pollock, Moore, Kreisler, \& Burch}]{gershman_mms_2022-1}
---. 2022{\natexlab{b}}, {MMS} 1 {Fast} {Plasma} {Investigation}, {Dual} {Ion} {Spectrometer} ({FPI}, {DIS}) {Instrument} {Distributions}, {Level} 2 ({L2}), {Burst} {Mode}, 0.15 s {Data},  NASA Space Physics Data Facility, \dodoi{10.48322/DQ1Y-NF73}

\bibitem[{Gershman {et~al.}(2022{\natexlab{c}})Gershman, Giles, Pollock, Moore, Kreisler, \& Burch}]{gershman_mms_2022-2}
---. 2022{\natexlab{c}}, {MMS} 1 {Fast} {Plasma} {Investigation}, {Dual} {Electron} {Spectrometer} ({FPI}, {DES}) {Instrument} {Distributions}, {Level} 2 ({L2}), {Burst} {Mode}, 30 ms {Data},  NASA Space Physics Data Facility, \dodoi{10.48322/1RPT-0W56}

\bibitem[{Gomez {et~al.}(2019)Gomez, Fuselier, Mukherjee, Gonzalez, Burch, Strangeway, \& Starkey}]{gomez_extramagnetospheric_2019}
Gomez, R.~G., Fuselier, S.~A., Mukherjee, J., {et~al.} 2019, Journal of Geophysical Research: Space Physics, 124, 1509, \dodoi{10.1029/2018JA025392}

\bibitem[{Gosling {et~al.}(1980)Gosling, Asbridge, Bame, Feldman, \& Zwickl}]{gosling_observations_1980}
Gosling, J., Asbridge, J., Bame, S., Feldman, W., \& Zwickl, R. 1980, Journal of Geophysical Research: Space Physics, 85, 3431, \dodoi{10.1029/JA085iA07p03431}

\bibitem[{Graham {et~al.}(2024)Graham, Khotyaintsev, Dimmock, Lalti, Boldú, Tigik, \& Fuselier}]{graham_ion_2024}
Graham, D.~B., Khotyaintsev, Y.~V., Dimmock, A.~P., {et~al.} 2024, Journal of Geophysical Research: Space Physics, 129, e2023JA032296, \dodoi{10.1029/2023JA032296}

\bibitem[{Kennel {et~al.}(1985)Kennel, Edmiston, \& Hada}]{kennel_quarter_1985}
Kennel, C.~F., Edmiston, J.~P., \& Hada, T. 1985, in Collisionless {Shocks} in the {Heliosphere}: {A} {Tutorial} {Review} (American Geophysical Union (AGU)), 1--36, \dodoi{https://doi.org/10.1029/GM034p0001}

\bibitem[{Koval \& Szabo(2021)}]{koval_wind_2021}
Koval, A., \& Szabo, A. 2021, Wind {Magnetic} {Field} {Investigation} ({MFI}) {Data} at full resolution,  NASA Space Physics Data Facility, \dodoi{10.48322/0V0H-DF27}

\bibitem[{Krasnoselskikh {et~al.}(2013)Krasnoselskikh, Balikhin, Walker, Schwartz, Sundkvist, Lobzin, Gedalin, Bale, Mozer, Soucek, Hobara, \& Comisel}]{krasnoselskikh_dynamic_2013}
Krasnoselskikh, V., Balikhin, M., Walker, S.~N., {et~al.} 2013, Space Science Reviews, 178, 535, \dodoi{10.1007/s11214-013-9972-y}

\bibitem[{Kropotina {et~al.}(2016)Kropotina, Bykov, Krasil’shchikov, \& Levenfish}]{kropotina_relaxation_2016}
Kropotina, Y.~A., Bykov, A.~M., Krasil’shchikov, A.~M., \& Levenfish, K.~P. 2016, Technical Physics, 61, 517, \dodoi{10.1134/S1063784216040149}

\bibitem[{Lee \& Wu(2000)}]{lee_heating_2000}
Lee, L.~C., \& Wu, B.~H. 2000, The Astrophysical Journal, 535, 1014, \dodoi{10.1086/308879}

\bibitem[{Lehmann \& Wardle(2016)}]{lehmann_signatures_2016}
Lehmann, A., \& Wardle, M. 2016, Monthly Notices of the Royal Astronomical Society, 455, 2066, \dodoi{10.1093/mnras/stv2311}

\bibitem[{Lei {et~al.}(2024)Lei, Zhou, Pang, Zhong, \& Deng}]{lei_quantified_2024}
Lei, G.~Y., Zhou, M., Pang, Y., Zhong, Z.~H., \& Deng, X.~H. 2024, The Astrophysical Journal, 964, 156, \dodoi{10.3847/1538-4357/ad2faf}

\bibitem[{Lepping {et~al.}(1995)Lepping, Acũna, Burlaga, Farrell, Slavin, Schatten, Mariani, Ness, Neubauer, Whang, Byrnes, Kennon, Panetta, Scheifele, \& Worley}]{lepping_wind_1995}
Lepping, R.~P., Acũna, M.~H., Burlaga, L.~F., {et~al.} 1995, Space Science Reviews, 71, 207, \dodoi{10.1007/BF00751330}

\bibitem[{Lin {et~al.}(1995)Lin, Anderson, Ashford, Carlson, Curtis, Ergun, Larson, McFadden, McCarthy, Parks, R�me, Bosqued, Coutelier, Cotin, D'Uston, Wenzel, Sanderson, Henrion, Ronnet, \& Paschmann}]{lin_three-dimensional_1995}
Lin, R.~P., Anderson, K.~A., Ashford, S., {et~al.} 1995, Space Science Reviews, 71, 125, \dodoi{10.1007/BF00751328}

\bibitem[{Madanian {et~al.}(2024)Madanian, Gingell, Chen, \& Monyek}]{madanian_drivers_2024}
Madanian, H., Gingell, I., Chen, L.-J., \& Monyek, E. 2024, The Astrophysical Journal Letters, 965, L12, \dodoi{10.3847/2041-8213/ad3073}

\bibitem[{Madanian {et~al.}(2022)Madanian, Liu, Phan, Trattner, Karlsson, \& Liemohn}]{madanian_asymmetric_2022}
Madanian, H., Liu, T.~Z., Phan, T.~D., {et~al.} 2022, Journal of Geophysical Research: Space Physics, 127, \dodoi{10.1029/2021JA030079}

\bibitem[{Madanian {et~al.}(2021{\natexlab{a}})Madanian, Schwartz, Fuselier, Burgess, Turner, Chen, Desai, \& Starkey}]{madanian_direct_2021}
Madanian, H., Schwartz, S.~J., Fuselier, S.~A., {et~al.} 2021{\natexlab{a}}, The Astrophysical Journal Letters, 915, L19, \dodoi{10.3847/2041-8213/ac0aee}

\bibitem[{Madanian {et~al.}(2021{\natexlab{b}})Madanian, Desai, Schwartz, Wilson, Fuselier, Burch, Contel, Turner, Ogasawara, Brosius, Russell, Ergun, Ahmadi, Gershman, \& Lindqvist}]{madanian_dynamics_2021}
Madanian, H., Desai, M.~I., Schwartz, S.~J., {et~al.} 2021{\natexlab{b}}, The Astrophysical Journal, 908, 40, \dodoi{10.3847/1538-4357/abcb88}

\bibitem[{McKean {et~al.}(1996)McKean, {Omidi, N.}, \& {Krauss-Varban, D.}}]{mckean_magnetosheath_1996}
McKean, M.~E., {Omidi, N.}, \& {Krauss-Varban, D.} 1996, Journal of Geophysical Research: Space Physics, 101, 20013, \dodoi{10.1029/96JA01461}

\bibitem[{Pollock {et~al.}(2016)Pollock, Moore, Jacques, Burch, Gliese, Saito, Omoto, Avanov, Barrie, Coffey, Dorelli, Gershman, Giles, Rosnack, Salo, Yokota, Adrian, Aoustin, Auletti, Aung, Bigio, Cao, Chandler, Chornay, Christian, Clark, Collinson, Corris, De Los Santos, Devlin, Diaz, Dickerson, Dickson, Diekmann, Diggs, Duncan, Figueroa-Vinas, Firman, Freeman, Galassi, Garcia, Goodhart, Guererro, Hageman, Hanley, Hemminger, Holland, Hutchins, James, Jones, Kreisler, Kujawski, Lavu, Lobell, LeCompte, Lukemire, MacDonald, Mariano, Mukai, Narayanan, Nguyan, Onizuka, Paterson, Persyn, Piepgrass, Cheney, Rager, Raghuram, Ramil, Reichenthal, Rodriguez, Rouzaud, Rucker, Saito, Samara, Sauvaud, Schuster, Shappirio, Shelton, Sher, Smith, Smith, Smith, Steinfeld, Szymkiewicz, Tanimoto, Taylor, Tucker, Tull, Uhl, Vloet, Walpole, Weidner, White, Winkert, Yeh, \& Zeuch}]{pollock_fast_2016}
Pollock, C., Moore, T., Jacques, A., {et~al.} 2016, Space Science Reviews, 199, 331, \dodoi{10.1007/s11214-016-0245-4}

\bibitem[{Preisser {et~al.}(2020)Preisser, Blanco‐Cano, Trotta, Burgess, \& Kajdič}]{preisser_influence_2020}
Preisser, L., Blanco‐Cano, X., Trotta, D., Burgess, D., \& Kajdič, P. 2020, Journal of Geophysical Research: Space Physics, 125, \dodoi{10.1029/2019JA027442}

\bibitem[{Russell {et~al.}(2022{\natexlab{a}})Russell, Magnes, Wei, Bromund, Plaschke, Fischer, Strangeway, Leinweber, Eichelberger, Huang, Le, \& Burch}]{russell_mms_2022}
Russell, C.~T., Magnes, W., Wei, H., {et~al.} 2022{\natexlab{a}}, {MMS} 2 {Flux} {Gate} {Magnetometer} ({FGM}) {DC} {Magnetic} {Field}, {Level} 2 ({L2}), {Survey} {Mode}, 8 or 16 {Sample}/s, v4/5 {Data},  NASA Space Physics Data Facility, \dodoi{10.48322/XZ3N-G079}

\bibitem[{Russell {et~al.}(2022{\natexlab{b}})Russell, Magnes, Wei, Bromund, Plaschke, Fischer, Strangeway, Leinweber, Eichelberger, Huang, Le, \& Burch}]{russell_mms_2022-1}
---. 2022{\natexlab{b}}, {MMS} 2 {Flux} {Gate} {Magnetometer} ({FGM}) {DC} {Magnetic} {Field}, {Level} 2 ({L2}), {Burst} {Mode}, 128 {Sample}/s, v4/5 {Data},  NASA Space Physics Data Facility, \dodoi{10.48322/GGX2-ZG64}

\bibitem[{Scholer \& Burgess(2007)}]{scholer_whistler_2007}
Scholer, M., \& Burgess, D. 2007, Physics of Plasmas, 14, 072103, \dodoi{10.1063/1.2748391}

\bibitem[{Schwartz(1998)}]{schwartz_analysis_1998}
Schwartz, S.~J. 1998, in {ISSI} {Scientific} {Reports} {Series}, Vol.~1, Analysis methods for multi-spacecraft data (ESA), 249--270

\bibitem[{Schwartz \& Burgess(1991)}]{schwartz_quasi-parallel_1991}
Schwartz, S.~J., \& Burgess, D. 1991, Geophysical Research Letters, 18, 373, \dodoi{10.1029/91GL00138}

\bibitem[{Schwartz {et~al.}(2022)Schwartz, Goodrich, Wilson, Turner, Trattner, Kucharek, Gingell, Fuselier, Cohen, Madanian, Ergun, Gershman, \& Strangeway}]{schwartz_energy_2022}
Schwartz, S.~J., Goodrich, K.~A., Wilson, L.~B., {et~al.} 2022, Journal of Geophysical Research: Space Physics, 127, e2022JA030637, \dodoi{10.1029/2022JA030637}

\bibitem[{Schwenn {et~al.}(1980)Schwenn, Rosenbauer, \& Mühlhäuser}]{schwenn_singlyionized_1980}
Schwenn, R., Rosenbauer, H., \& Mühlhäuser, K. 1980, Geophysical Research Letters, 7, 201, \dodoi{10.1029/GL007i003p00201}

\bibitem[{Sckopke(1995)}]{sckopke_ion_1995}
Sckopke, N. 1995, Advances in Space Research, 15, 261, \dodoi{10.1016/0273-1177(94)00106-B}

\bibitem[{Skoug {et~al.}(1999)Skoug, Bame, Feldman, Gosling, McComas, Steinberg, Tokar, Riley, Burlaga, Ness, \& Smith}]{skoug_prolonged_1999}
Skoug, R.~M., Bame, S.~J., Feldman, W.~C., {et~al.} 1999, Geophysical Research Letters, 26, 161, \dodoi{10.1029/1998GL900207}

\bibitem[{Starkey {et~al.}(2020)Starkey, Fuselier, Desai, Schwartz, Gomez, Mukherjee, Cohen, \& Russell}]{starkey_mms_2020}
Starkey, M., Fuselier, S.~A., Desai, M.~I., {et~al.} 2020, The Astrophysical Journal, 897, 6, \dodoi{10.3847/1538-4357/ab960c}

\bibitem[{Starkey {et~al.}(2019)Starkey, Fuselier, Desai, Burch, Gomez, Mukherjee, Russell, Lai, \& Schwartz}]{starkey_acceleration_2019}
Starkey, M.~J., Fuselier, S.~A., Desai, M.~I., {et~al.} 2019, Geophysical Research Letters, 46, 10735, \dodoi{10.1029/2019GL084198}

\bibitem[{Thomsen {et~al.}(1985)Thomsen, Gosling, Bame, \& Mellott}]{thomsen_ion_1985}
Thomsen, M.~F., Gosling, J.~T., Bame, S.~J., \& Mellott, M.~M. 1985, Journal of Geophysical Research: Space Physics, 90, 137, \dodoi{10.1029/JA090iA01p00137}

\bibitem[{Tidman \& Northrop(1968)}]{tidman_emission_1968}
Tidman, D.~A., \& Northrop, T.~G. 1968, Journal of Geophysical Research, 73, 1543, \dodoi{10.1029/JA073i005p01543}

\bibitem[{Treumann(2009)}]{treumann_fundamentals_2009}
Treumann, R.~A. 2009, The Astronomy and Astrophysics Review, 17, 409, \dodoi{10.1007/s00159-009-0024-2}

\bibitem[{Wang {et~al.}(2018)Wang, Feng, \& Zhao}]{wang_cold_2018}
Wang, J., Feng, H., \& Zhao, G. 2018, Astronomy \& Astrophysics, 616, A41, \dodoi{10.1051/0004-6361/201731807}

\bibitem[{Wilkinson(1991)}]{wilkinson_ion_1991}
Wilkinson, W.~P. 1991, Journal of Geophysical Research: Space Physics, 96, 17675, \dodoi{10.1029/91JA01646}

\bibitem[{Wilson {et~al.}(2021{\natexlab{a}})Wilson, Chen, \& Roytershteyn}]{wilson_discrepancy_2021}
Wilson, L.~B., Chen, L.-J., \& Roytershteyn, V. 2021{\natexlab{a}}, Frontiers in Astronomy and Space Sciences, 7, 592634, \dodoi{10.3389/fspas.2020.592634}

\bibitem[{Wilson {et~al.}(2021{\natexlab{b}})Wilson, Brosius, Gopalswamy, Nieves‐Chinchilla, Szabo, Hurley, Phan, Kasper, Lugaz, Richardson, Chen, Verscharen, Wicks, \& TenBarge}]{wilson_quarter_2021}
Wilson, L.~B., Brosius, A.~L., Gopalswamy, N., {et~al.} 2021{\natexlab{b}}, Reviews of Geophysics, 59, \dodoi{10.1029/2020RG000714}

\bibitem[{Wilson~III {et~al.}(2012)Wilson~III, Koval, Szabo, Breneman, Cattell, Goetz, Kellogg, Kersten, Kasper, Maruca, \& Pulupa}]{wilson_observations_2012}
Wilson~III, L.~B., Koval, A., Szabo, A., {et~al.} 2012, Geophysical Research Letters, 39, L08109, \dodoi{10.1029/2012GL051581}

\bibitem[{Wimmer-Schweingruber {et~al.}(2006)Wimmer-Schweingruber, Crooker, Balogh, Bothmer, Forsyth, Gazis, Gosling, Horbury, Kilchenmann, Richardson, Richardson, Riley, Rodriguez, Steiger, Wurz, \& Zurbuchen}]{wimmer-schweingruber_understanding_2006}
Wimmer-Schweingruber, R.~F., Crooker, N.~U., Balogh, A., {et~al.} 2006, Space Science Reviews, 123, 177, \dodoi{10.1007/s11214-006-9017-x}

\bibitem[{Wu {et~al.}(2009)Wu, Winske, Gary, Schwadron, \& Lee}]{wu_energy_2009}
Wu, P., Winske, D., Gary, S.~P., Schwadron, N.~A., \& Lee, M.~A. 2009, Journal of Geophysical Research: Space Physics, 114, 2009JA014240, \dodoi{10.1029/2009JA014240}

\bibitem[{Young {et~al.}(2016)Young, Burch, Gomez, De~Los~Santos, Miller, Wilson, Paschalidis, Fuselier, Pickens, Hertzberg, Pollock, Scherrer, Wood, Donald, Aaron, Furman, George, Gurnee, Hourani, Jacques, Johnson, Orr, Pan, Persyn, Pope, Roberts, Stokes, Trattner, \& Webster}]{young_hot_2016}
Young, D.~T., Burch, J.~L., Gomez, R.~G., {et~al.} 2016, Space Science Reviews, 199, 407, \dodoi{10.1007/s11214-014-0119-6}

\bibitem[{Zou {et~al.}(2024)Zou, Gjerloev, Ohtani, Friel, Liang, Lyons, Shen, Liu, Chen, Ferdousi, Chartier, Vines, \& Waters}]{zou_extreme_2024}
Zou, Y., Gjerloev, J.~W., Ohtani, S., {et~al.} 2024, AGU Advances, 5, e2023AV001101, \dodoi{10.1029/2023AV001101}

\end{thebibliography}
\bibliographystyle{aasjournal}

\end{document}